\documentclass[12pt]{iopart}
\usepackage{graphicx}
\usepackage{iopams}
\usepackage{graphicx,braket,xcolor,subfigure,upgreek}
\usepackage[colorlinks, linkcolor=blue, citecolor=blue, urlcolor=blue, breaklinks=true]{hyperref}
\usepackage{microtype}
\usepackage{bbm}
\usepackage{color}
\usepackage[english]{babel}

\newcommand{\frefa}[1]{Fig.~\ref{#1}}

\begin{document}
	
\title[Quantum Improvement]{Unraveling the origin of higher success probabilities in quantum annealing versus semi-classical annealing}
	
\author{Elias Starchl and Helmut Ritsch}
\ead{Elias.Starchl@uibk.ac.at}
\address{Institut f\"ur Theoretische Physik, Universit\"at Innsbruck, Technikerstr. 21a, A-6020 Innsbruck, Austria}
\submitto{JPhysB}
\date{\today}

\begin{abstract}
Quantum annealing aims at finding optimal solutions to complex optimization problems using a suitable quantum many body Hamiltonian encoding the solution in its ground state. To find the solution one typically evolves the ground state of a soluble, simple initial Hamiltonian adiabatically to the ground state of the designated final Hamiltonian. Here we explore whether and when a full quantum representation of the dynamics leads to higher probability to end up in the desired ground when compared to a classical mean field approximation. As simple but nontrivial example we target the ground state of interacting bosons trapped in a tight binding lattice with small local defect by turning on long range interactions. Already two atoms in four sites interacting via two cavity modes prove complex enough to exhibit significant differences between the full quantum model and a mean field approximation for the cavity fields mediating the interactions. We find a large parameter region of highly successful quantum annealing, where the semi-classical approach largely fails. Here we see strong evidence for the importance of entanglement to end close to the optimal solution. The quantum model also reduces the minimal time for a high target occupation probability. Surprisingly, in contrast to naive expectations that enlarging the Hilbert space is beneficial, different numerical cut-offs of the Hilbert space reveal an improved performance for lower cut-offs, i.e. an nonphysical reduced Hilbert space, for short simulation times. Hence a less faithful representation of the full quantum dynamics sometimes creates a higher numerical success probability in even shorter time. However, a sufficiently high cut-off proves relevant to obtain near perfect fidelity for long simulations times in a single run. Overall our results exhibit a clear improvement to find the optimal solution based on a quantum model versus simulations based on a classical field approximation.   
\end{abstract}


\maketitle

\section{Introduction}
Quantum computation (QC) has been proposed already in the early 80's when Paul Benioff introduced a quantum Turing machine \cite{benioff1980}, Feynman the universal quantum simulator \cite{feynman1982} and David Deutsch a universal circuit quantum computer \cite{deutsch1985}. In the past years its technical implementation made immense progress partly driven by huge tech companies like Google and IBM. Most recently, strong claims have been put forward that quantum machines have solved sample problems not accessible to the best classical computers to date \cite{dowling2002}.
\par
As full scale programmable quantum computers are technically extremely challenging to implement, many current efforts are devoted to adiabatic quantum simulation. Here one aims to prepare the ground state of a many-body Hamiltonian, which encodes the solution to a classical optimization problem in tailored interactions\cite{das2005}. In a typical approach one starts from the ground state of a simple initial many body Hamiltonian and evolves this state to the desired ground state via a slow ramp up of the particle interactions towards the target Hamiltonian. 

Notably, one of the earliest work in laying the foundation of quantum annealing, was done in 1989 \cite{ray1989}, showing that quantum fluctuations can increase the ergodicity in a spin-glass model, by tunneling between "trapping" minima, separated by narrow potential barriers. The idea of using tailored interactions of a many-body Hamiltonian, originally suggested as quantum stochastic optimization~\cite{apolloni1989}, was later renamed quantum annealing (QA) \cite{apolloni1990}. Nearly a decade later, in 1998, the present form of QA was introduced by considering adjustable quantum fluctuation instead of thermal ones \cite{kadowaki1998}. This basic idea, of integrating an artificial kinetic term, introducing quantum fluctuations in a controlled way, still remains a powerful tool and also provides the principal technique for our work.
The theoretical foundation for this kind of adiabatic evolution is given by the adiabatic theorem \cite{born1928} which states, that the systems stays in its ground state, if the Hamiltonian is changed slowly on the time scale set by the energy gap from the ground to excited states.
\par
The QA protocol promises practically useful results even for small scale quantum simulators with less than a hundred q-bits. Experimental examples are trapped-ion simulators \cite{britton2012}, superconducting qubits \cite{bunyk2014} or ultra cold atoms in optical lattices \cite{jotzu2014}. A commercially available superconducting quantum annealer was launched already in 2011 by D-Wave \cite{johnson2011} and is continuously enlarged and improved. However, it is still highly debated, if quantum annealing devices offer an actual speedup over classical devices\cite{hauke2020}. In several cases, optimizing algorithms on classical computers lead to a comparable or even better performance than the quantum machine at current size and speed\cite{heim2015}. Interestingly, in several cases quantum inspired algorithms\cite{arrazola2019}, i.e. the efficient numerical solution of the Schrödinger equation, proved to be the fastest available classical computer based method to solve the problem.
\par
Let introduce here some central early papers and reviews on this subject. With a main focus on quantum annealing and analog quantum computation the review by Das et al. in 2008 \cite{das2008} gives an introduction to the initial progress of the subject matter and a clear picture of the fundamental physical properties and mechanisms of QA. Further, a more recent review by Hauke et al. in 2020 \cite{hauke2020} focuses more on the present state of QA and new pathways and possibilities in theory and experiment. Finally, the review from Albash et al. in 2018 \cite{albash2018a} gives a more comprehensive perspective on quantum adiabatic computation in general, its theoretical development, algorithms and computational complexity.
\par
A generic example to study a possible advantage of quantum annealing versus classical computation was put forward in a recent proposal for a quantum annealer using ultra-cold atoms in an optical lattice within a cavity\cite{torggler2017}. Here the cavity modes mediate controllable all to all interactions between the atoms in the lattice. As key example an implementation of the quantum annealing process is predicted to generate solutions of the N-queens completion problem\cite{letavec2002,torggler2019}, which is known to be NP-complete \cite{gent2017}. Interestingly one finds that for the optimal quantum implementation of the problem one only needs a linear increase in resources (Q-bits and lasers) with problem size. Thus the record size classical solution of N=23 found so far\cite{das2005} seems well in reach of current optical lattice experiments. Numerically, corresponding solutions of the time dependent Schrödinger equation are hardly possible for more than 8 q-bits. Unfortunately the precise system size scaling of the minimal gap of the corresponding Hamiltonian could neither be analytically calculated nor numerically reliably estimated. Hence the confirmation of a corresponding quantum advantage is left to experiment. 

Nevertheless the numerical studies of the above model reveal some interesting properties, which indirectly could shed some extra light on the beneficial role of quantum mechanics in the annealing process. Implementing different approximate numerical models to simulate the time-dependent numerical annealing dynamics, one finds, that the most reliable algorithm to find the ground with high probability required a full quantum description of the cavity field dynamics and the atomic motion. A so-called semi-classical mean-field approach, where the cavity fields are approximated by coherent fields failed to find the correct solution with a useful probability. As solutions for nontrivial problem sizes (i.e. chess board sizes N$>$8)  are numerically very complex, the detailed physical origin behind this form of a quantum improvement could not be studied in detail.

Note that here the notion of quantum improvement or advantage appears in a bit unusual and more literal sense. We simply tie it to the observation that a full quantum description of the cavity fields in the numerical annealing simulation yields a significantly higher probability to end up in the desired final ground state than using a classical description of the light modes. Note that the classical field dynamics is numerically much simpler to implement and thus, in principle, the solution can be computed faster. This could compensate the lower success probability for a single run but for our line of argumentation towards a quantum improvement we compare success probabilities for equal simulation times. 
\par
The idea of this work here is to take one step back and devise a related, but sufficiently simplified toy model to unravel the origin of this special sort of quantum advantage or improvement. The challenge here is to allow for enough model complexity to clearly see the improved performance of a full quantum treatment but still keep the model numerically tractable and conceptually clear enough to analyze the underlying dynamics in sufficient detail.

As we will exhibit below, the study of self-ordering in a finite one-dimensional optical lattice with long range interactions mediated by two independent cavity modes satisfies both requirements. When we add a small defect in the lattice, the semi-classical approach largely fails to find the correct ground state for a wide and well defined range of parameters, whereas the full quantum dynamics still reliably arrives at the correct solution. Note that finding the ground state in a system with very small disorder also is a complex problem for larger instances\cite{dagotto2005complexity}. Hence we now have found a suitable test ground to study the physical origin and conditions for the superior performance of a quantum simulator, which in much more complex configurations could be the central ingredient to obtain a quantum advantage.
\par
This work is organized as follows.  In section \ref{sec:model} we present our quantum gas cavity QED Bose-Hubbard (BH) Hamiltonian\cite{mekhov2012} and shortly review the basic concepts of quantum annealing based on the the adiabatic approximation. In the next chapter \ref{sec:analysis quantum} we study the conditions and parameters needed for successful quantum annealing and the role of entanglement in this process. This is followed by a comparison of the quantum dynamics to the semi-classical mean-field description \ref{sec:semi-class}. In a final chapter we present the somewhat surprising and counter-intuitive dependence of the optimization efficiency on the numerical photon cut-off, where sometimes a numerically less precise implementation of the quantum time evolution improves the annealing efficiency as shown in section \ref{sec:cut-off}.

\begin{figure}[h!]
	\centering
	\includegraphics[scale=0.3]{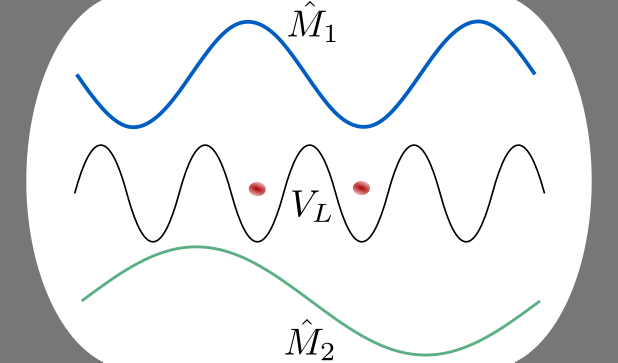}
	\caption{Scheme of the model system of two bosonic particles moving in a four site lattice with long range interactions mediated by light scattering into two independent cavity modes.}
	\label{fig:model}
\end{figure}

\section{Model Hamiltonian and Approximations}\label{sec:model}
Let us consider a one dimensional lattice described by a tight binding BH model with four sites with periodic boundary conditions filled by two interacting bosons. The particles interact on site via collisions and we introduce tunable non-local interactions via collective light scattering to two independent cavity field modes \cite{maschler2008,mekhov2012} as depicted in \frefa{fig:model}. The open system dynamics of the cavity fields mediating the interaction can be described via quantum oscillators or it can be approximated by classical fields, as c-number amplitudes. As the pump lasers illuminating the atoms are described by coherent fields and working in the weak excitation regime, the latter can be expected to give an approximation for the full quantum dynamics \cite{maschler2008}. As we assume the cavity damping rates much faster than the atomic dynamics, an adiabatic elimination of the field dynamics should well approximate the induced all to all coupling.\\

\noindent
{\bf Full quantum dynamics:} Let us start by introducing the full quantum model for the coupled dynamics of atoms and field modes. For all levels of description of the atomic dynamics we will use the same BH-type particle Hamiltonian

\begin{equation}
 H_{kin}=J\sum_{k_{PBC}}(b_k^\dagger b_{k+1} + h.c.) \;\mbox{and}\; U_{int} =\frac{U}{2} \sum_k n_k(n_k - 1),
\end{equation}
where $k\in \left\{1,2,3,4\right\}$ and $k_{PBC}$ denotes the periodic boundary conditions. Of course, it would be interesting to also consider the case where we replace this by a classical Brownian motion model \cite{torggler2014} but this is beyond the aim of this work.  

The interactions between different lattice sites are mediated by cavity modes. As discussed in recent publications \cite{keller2018,kramer2014,torggler2017}, introducing two independent field modes with different wave-vectors already creates surprising complexity into this model. Hence we can expect generic quantum effects play a vital role. Assuming that there is no scattering between the modes, i.e. a big enough frequency spacing, we can simply add the interaction terms induced independently by each mode. Experimentally this simply requires adding an extra pump laser with a mode specific detuning $\Delta_c^m = \omega_p^m - \omega_c^m$. To keep it simple enough we will use two modes, where one has twice the wavelength of the other. In this case the local interaction strengths are periodic with equal strength and simply amount to site dependent sign changes.   

Incorporating these symmetries, we can write the long range interaction Hamiltonian as

\begin{equation}\label{Hint}
	H_{int} = \tilde{J}\big(\hat{M}_1 (a_1+a_1^\dagger) + \hat{M}_2 (a_2+a_2^\dagger)\big),
\end{equation}
where we introduced the overall scattering amplitude $\tilde{J}$ and the effective scattering operators determining the light scattering amplitudes into the modes, which explicitly read:

\begin{eqnarray} 
\hat{M}_1 &= (-\hat{n}_1 + \hat{n}_2 - \hat{n}_3 + \hat{n}_4) ,\; \\  \hat{M}_2 &= (+\hat{n}_1 + \hat{n}_2 - \hat{n}_3 - \hat{n}_4).
\end{eqnarray} 

In our present case the pump strength $\tilde{J}$ is equal for all modes and lattice sites. The cavity-pump detunings $\Delta$ are tuned to equal by adjusting the pump frequencies $\omega_p^1$ and $\omega_p^2$, therefore we can write the cavity radiation Hamiltonian as

\begin{equation}
H_{rad}= -\Delta (a_1^\dagger a_1 + a_2^\dagger a_2). \label{Hrad}
\end{equation}

The complete effective Hamiltonian finally reads

\begin{eqnarray}
	H &= J\sum_{k_{PBC}}(b_k^\dagger b_{k+1} + h.c.)  +\frac{U}{2} \sum_k n_k(n_k - 1)  \nonumber  \\
         -&\Delta (a_1^\dagger a_1 + a_2^\dagger a_2) +\tilde{J}\big(\hat{M}_1 (a_1+a_1^\dagger) 
     + \hat{M}_2 (a_2+a_2^\dagger)\big). \label{eq:H}
\end{eqnarray} 

As we deal with an open system, in principle we have to include dissipation via the cavity mirrors and spontaneous emission of photons. Assuming operation far from atomic resonance, we neglect the latter and incorporate only cavity loss $\kappa$ via a standard quantum optics master equation:

\begin{eqnarray}
	\dot{\rho} &= -\frac{i}{\hbar}\left[H,\rho\right] + \mathcal{L}[\rho], \\
\mathcal{L}[\rho] &= \sum_{m\in \left\{1,2\right\}}\kappa_m (2a_m\rho a_m^\dagger - a_m^\dagger a_m \rho - \rho a_m^\dagger a_m),
\end{eqnarray} 
where $\mathcal{L}$ is a standard damping Liouvillean \cite{breuer2002}. \\

{\bf Adiabatic elimination and semi-classical description of modes:} Our central aim in this work is to study the role of {\sl quantumness} in simulated annealing. In addition to the full quantum dynamics we thus will introduce two additional approximate descriptions, which differ in the description of the cavity mode as quantum fields or as classical fields. In both cases, for direct comparability we assume adiabatic following of the field modes to the atomic state, which allows for adiabatic elimination in the mode dynamics. In the first case we treat the field mode amplitudes as quantum operators governed by their Heisenberg equations, while in the second case they are described by classical coherent fields with c-number amplitudes governed by Maxwell's equations. Note that in the latter case no atom-field entanglement can build up.     
   
Using (\ref{Hint}) and (\ref{Hrad}) the Heisenberg equations for the field amplitude operators $\dot{a}_m = \frac{i}{\hbar}[H_{rad} + H_{int},a_m]$ read:

\begin{equation}\label{adot}
	\dot{a}_m = i \Delta a_m -i\tilde{J} \hat{M}_m  - \kappa_m a_m.
\end{equation}

We can construct an effective particle Hamiltonian by adiabatic elimination of the fields (\ref{adot}), which is valid if $\kappa_m$ and $\Delta$ are larger than the timescale of atomic motion $J/\hbar \ll |\Delta + i\kappa|$. Hence the fields will closely follow atomic dynamics which formally yields:

\begin{equation}\label{eq:a_m}
	a_m = \frac{-i \tilde{J}}{\kappa_m - i \Delta}\hat{M}_m.
\end{equation}

We will now proceed in two ways to derive an effective particle interaction Hamiltonian:

First we can insert the full operator expression $a_m$ in the Hamiltonian (\ref{Hint}) and (\ref{Hrad}), such that the atom-light interaction reduces to an effective interaction of the particles, so that we get:

\begin{equation}\label{eq:Had}
	H_{ad} = \frac{\Delta \tilde{J}^2}{\kappa^2 + \Delta^2}\big(\hat{M}_1^2 + \hat{M}_2^2\big).
\end{equation}

In the second alternative approach we further approximate the field dynamics using a common mean field assumption, where the mode operators are represented by classical c-number amplitudes $\alpha_m(t)$ determined by the expectation values $ \langle a_m\rangle = \alpha_m(t) \in C $ of the corresponding operators:

\begin{equation}\label{eq:alpha}
	\alpha_m(t) = \frac{-i\tilde{J}}{\kappa_m - i\Delta} \langle\hat{M}_m\rangle. 
\end{equation}

Inserting these approximate expressions for the fields into the original atom-field Hamiltonian defines our semi classical interaction Hamiltonian 

\begin{equation}\label{eq:H-semi-class}
	H_{ad}^{sc}= \frac{\Delta \tilde{J}^2}{\kappa^2 + \Delta^2}\big( 2\hat{M}_1\langle\hat{M}_1\rangle - \hat{{I}}\langle\hat{M}_1\rangle^2 +   2\hat{M}_2\langle\hat{M}_2\rangle - \hat{{I}}\langle\hat{M}_2\rangle^2 \big).
\end{equation}

Here $\hat{{I}}$ denotes the identity operator. Note that semi-classical approximations in this form are a well proven tool in quantum optics, laser physics and quantum gas cavity QED to reduce the complexity of coupled atom field solutions. To solve for the corresponding Schrödinger dynamics one has to evaluate the expectation values of the mode operators self-consistently at each timestep.    

 As mentioned we restrict ourselves to two particles within our four lattice sites, i.e. half filling. Accordingly the particle Hilbert space dimension of the space is $3^4=81$. Note that our two approximate Hamiltonians (\ref{eq:H-semi-class}) and (\ref{eq:Had}) formally look very similar, but finally will lead to even quantitatively different annealing results in the important parameter regime. 

At this point our 4-site Hamiltonian with periodic boundary conditions still is translation invariant and thus has close to degenerate ground states. In order to create a well defined target ground state for the annealing process, we will introduce a certain amount of impurity, adding a small perturbation $H_d = -(1+ V) \hat{n_3}$ at the third lattice site changing the depth of the third potential well. This part can be incorporated in the mode operators $M_i$ which simply change to: 

\begin{equation} \label{M(V)}
\hat{M_1}= (-\hat{n}_1 + \hat{n}_2 - V \hat{n}_3 + \hat{n}_4),\\ 
\hat{M}_2 = (+\hat{n}_1 + \hat{n}_2 - V \hat{n}_3 - \hat{n}_4).
\end{equation} 

With $V > 1$ we lift the degeneracy in the ground state of the energy spectrum and enforce a unique lowest energy solution in form of a two particle occupation state of the third lattice site for large pump $\tilde{J}$.

Note, that we work in the regime of weak excitation in the modes, hence low photon numbers and large detuning, and we consider a damping rate $\kappa$ faster than the timescale of the atomic dynamics, resulting in close to adiabatic field dynamics. Thus, the photon number fluctuations are fast and essentially average out, resulting in a coherent interaction. This allows us to consider only trajectories with unitary evolution. In principal, decay should be further investigated but, for simplicity, is not pursued in this work. Therefore, whenever we refer to the dynamics of the \emph{full quantum description} we consider evolution in terms of \ref{eq:H}.

\section{Quantum annealing }\label{sec:analysis quantum}
In this section we analyze the detailed properties of the annealing procedure based on a full quantum atom and field description. We dynamically evolve the ground state of the initial Bose Hubbard Hamiltonian without the laser induced interactions $\tilde{J} = 0$ adiabatically towards the ground state of the full interacting Hamiltonian at large pumping $\tilde{J} \gg J$, including weak impurity. 

Here it is helpful to first look at the energy spectrum of the time dependent Hamiltonian as a function of the pump strength allowing to identify the minimum excitation gap as decisive parameter determining the quality of the annealing process. This gap limits the minimum time for an adiabatic ramp up procedure of the laser induced interactions. All energies and parameters here are given in units of the recoil energy $\hbar \omega_r = \hbar\; \frac{\hbar k^2}{2m}$, corresponding to the kinetic energy of an atom with mass $m$ with a momentum equal to the single photon momentum $\hbar k$, with $\omega_r$ the recoil frequency. 

\begin{figure}[h]
	\centering
	\includegraphics[width = 0.6\textwidth]{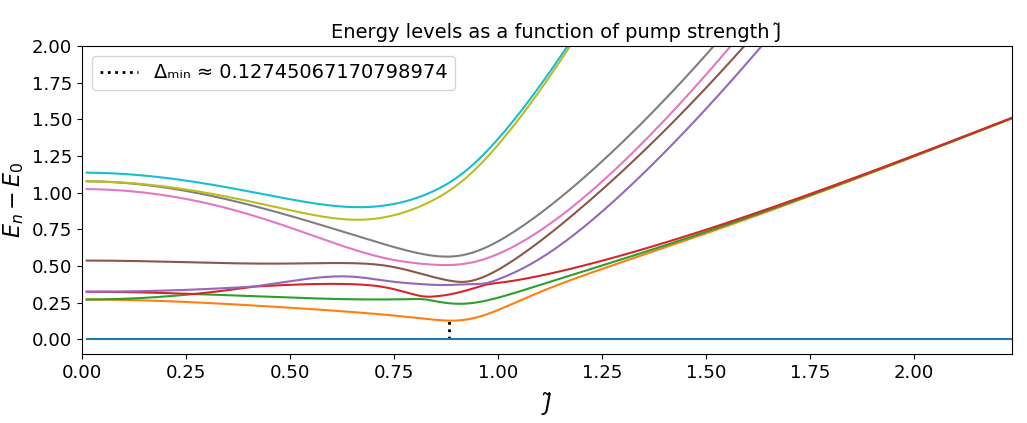}
	\caption{Energy spectrum of the full quantum atom-field  Hamiltonian (\ref{eq:H}) up to the 10th eigenstate. Energies are plotted relative to the ground state as a function of effective pump strength $\tilde{J}(s)$ with $s\in [0,1]$. The minimal energy gap $\Delta_{min}\approx0.1275$ between ground and first excited state is annotated by a dashed vertical line depicted. The chosen operating parameters are: $J=0.1$, $U=0.7$, $V=1.1$, $\tilde{J}^2=5=\Delta$.}
	\label{fig:energy-spectrum}
\end{figure}

In \frefa{fig:energy-spectrum} we plot the numerically determined lowest 10 eigenvalues $H(s)\ket{\psi_n(s)}=E_n(s)\ket{\psi_n(s)}$ relative to the ground state, i.e $E_n(s) - E_0(s)$, for different effective pump strength $\tilde{J(s)}$. Note that the parameter $s=t/t_f$ describes the time dependence of the pump strength, with the simulation time $t_f$ and $t\in[0,t_f]$. At this point it is important to note that we need to introduce a photon number cut-off (denoted by \textit{nc}), as the Hilbert space of a field mode (= harmonic oscillator) would have infinite dimension. For the following simulations we use a cut-off of $nc=3$ and further investigations will be presented in the final section. We see that for any chosen finite final impurity $V$ a gap remains open and takes on its minimum of about a recoil energy close to unit pump strength $\tilde{J} = 1$. 

\begin{figure}[h!]
	\centering
	\includegraphics[width = 0.66\textwidth]{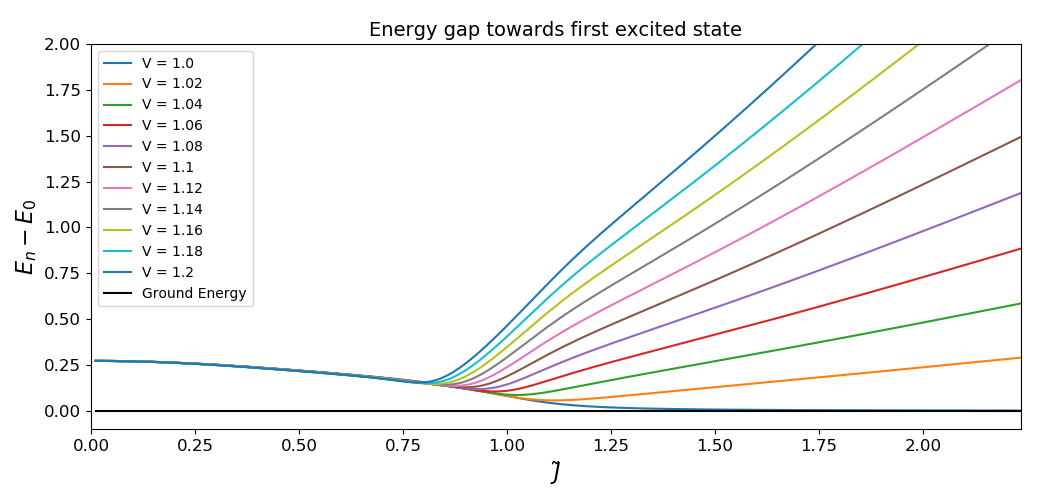}
	\caption{First excited state energy for different impurity well-depth $V$ relative to the ground state as a function of $\tilde{J}(s)$ with $s\in [0,1]$. An energy gap and a lifting of the ground state degeneracy at large pump only occurs when we introduce impurity with $V > 1$ increases. System parameters are chosen as: $J=0.1$, $U=0.7$, $V\in [1.0,1.2]$, $\tilde{J}^2=5=\Delta$.}
	\label{fig:energy-V}
\end{figure}

\begin{figure}[h!]
	\centering
	\includegraphics[width = 0.5\textwidth]{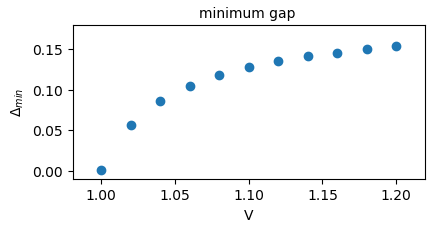}
	\caption{Minimum energy gap between ground and first excited state as a function of the impurity magnitude $V$ for: $J=0.1$, $U=0.7$, $V\in[1.0,1.2]$, $\tilde{J}^2=5=\Delta$}
	\label{fig:delta-V}
\end{figure}

The gap size depends on the impurity potential $V$ as depicted in \frefa{fig:energy-V} and \frefa{fig:delta-V}. As expected from symmetry arguments, without added impurity $V=1$ we get two almost degenerate ground states and thus a vanishing gap for large pump strength $\tilde{J}$.
Note that the minimum gap depends nonlinearly on $V$ as shown in \frefa{fig:delta-V}.

\begin{figure}[h!]
	\centering
	\includegraphics[width = 0.75\textwidth]{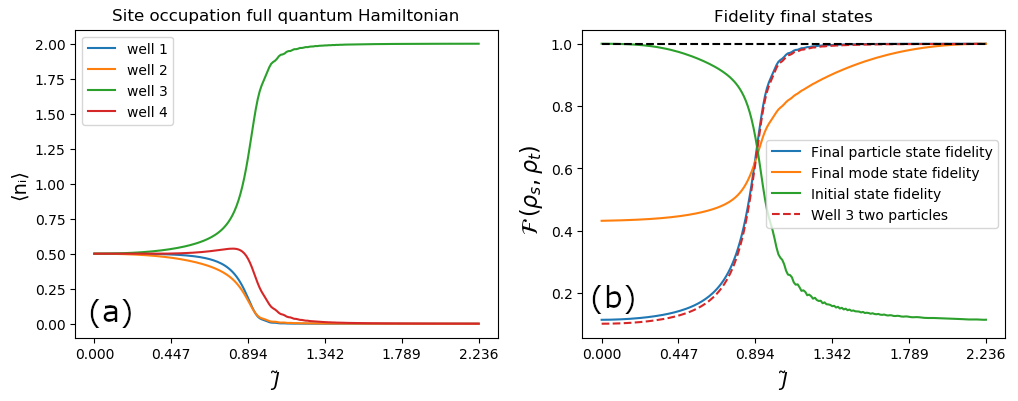}
	\caption{State evolution for the full quantum Hamiltonian during one adiabatic sweep. In the left graph (a) the expectation values of the lattice site occupations are plotted, while the right figure (b) exhibits the corresponding evolution of the overlaps (fidelity) between the evolving state and the target solution as well as the initial state. The dashed line gives the probability for two atoms in the third well as an alternative measure of the annealing success. Parameters are: $J=0.1$, $U=0.7$, $V=1.1$, $\tilde{J}^2=5=\Delta$.}
	\label{fig:Hf_evo}
\end{figure}

Let us now start to study the corresponding annealing process. In its simplest form we start the dynamics with the system ground state at zero pump $\tilde{J} = 0$ and linearly increase pump strength towards $\tilde{J} = \sqrt{5}$. For a sufficiently slow ramp time much longer than the recoil time we evolve this state via the corresponding Schrödinger equation with a time dependent Hamiltonian $H$. In \frefa{fig:Hf_evo} (a) we show the site occupation probabilities as a function of time for the example of a sufficiently slow sweep, such that the state evolves adiabatically. As expected, around $\tilde{J} \approx 1$ both atoms are transferred to the deeper third well with very high probability.

The efficiency of this process can be quantified by the probability to reach the ground state of the final Hamiltonian as a function of various simulation parameters. This is shown in \frefa{fig:Hf_evo} (b) together with probability for the two atom occupation probability of the third well. We clearly see that both probabilities almost reach unity at the end for slow enough annealing. Note that the overlap of the field component of the full quantum state with the adiabatic eigenstate is significantly lower during evolution. This shows that the field dynamics lags somewhat behind its adiabatic value without preventing very good annealing success at the end.    

\begin{figure}[h!]
	\centering 
\includegraphics[width = 0.3\textwidth]{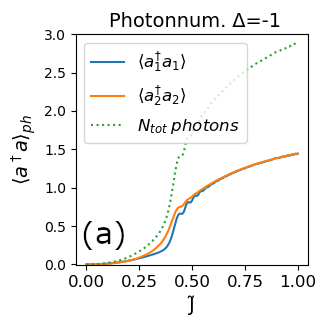}
\includegraphics[width = 0.3\textwidth5]{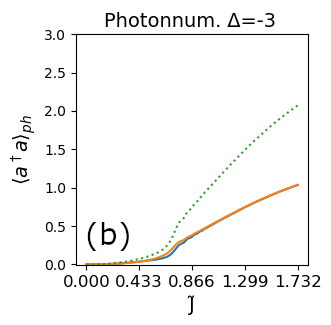}
\includegraphics[width = 0.3\textwidth5]{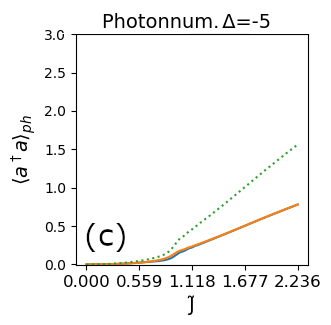}
	\caption{Evolution of the average cavity photon numbers in both modes and their sum as a function of the time-dependent pump-strength for three different choices of cavity detunings $\Delta = -1 $ (a), $\Delta = -3 $ (b) and $\Delta = -5 $ (c).The final pump amplitudes are increased as $\tilde{J}=\sqrt{-\Delta}$. The kink is directly connected to atomic ordering.}
	\label{fig:photons}
\end{figure}

The reordering of the atoms is accompanied and thus can be directly monitored by the buildup of the cavity field amplitudes as depicted in \frefa{fig:photons}. Here the time evolution of the photon numbers is depicted as a function of momentary pump strength $\tilde{J}(s)$. Increasing detuning and pump strength leads to lower photon numbers during the annealing process without much influence on the success. This operating regime allows to reduce the effective dissipation via cavity loss and also allows to use a lower photon cutoff in the numerical simulations saving computational time.\par

As mentioned above, adiabatic following requires a slow time evolution of the Hamiltonian. It is, of course, of vital interest here to find out, how fast one can progress without significantly lowering the success rate. In general one operates in the regime where the particle-particle interaction mediated by the field modes closely follows the particles state, i.e. the relaxation time scale of the modes is fast. However, as we shall see, instant following of the fields relative to the atomic dynamics, as mathematically assumed in the adiabatically eliminated quantum Hamiltonian, is not always optimal. To exhibit this, the success probability as a function of simulation time is depicted in the next two graphs, where we plot the final state overlap with the target state as a function of ramp time. Here we compare the full dynamic and the adiabatically eliminated case. 

\begin{figure}[h!]\centering
	\includegraphics[scale=0.4]{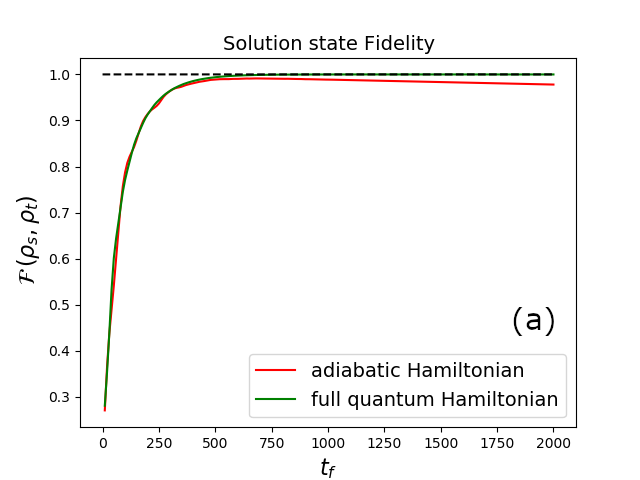} \includegraphics[scale=0.4]{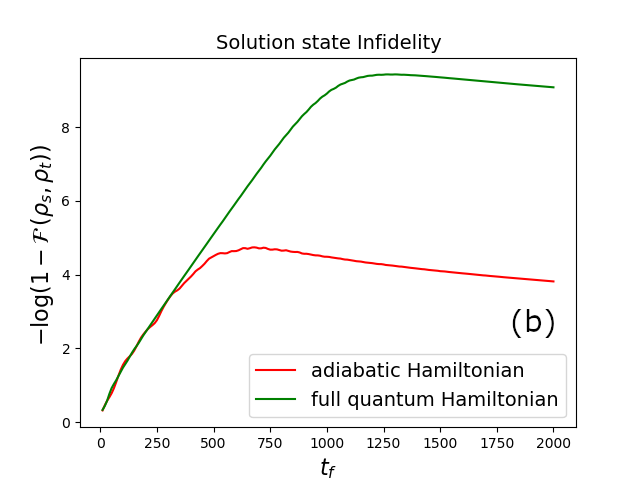}
	\caption{(a) The fidelity between the final state after adiabatic evolution and the desired target state as a function of simulation time. The green curve corresponds to the Hamiltonian with full dynamic quantum fields (\ref{eq:H}) while the red line represents the adiabatic elimination model (\ref{eq:Had}).
(b) The negative logarithm of the infidelity as a function of simulation time. Parameters are: $J=0.1$, $U=0.7$, $V=1.1$, $\tilde{J}^2=5=\Delta$ and $\kappa=1$.
 }
	\label{fig:nc3fid}
\end{figure}

As we see, for simulation times $\omega_r*t_f > 250 $ the success probability (fidelity) shown in \frefa{fig:nc3fid} (a) in both cases well surpasses $0.9$ for the full and the adiabatic Hamiltonian (\ref{eq:Had}). On this scale the two curves agree pretty well, although the adiabatic Hamiltonian effectively incorporates photon loss via the cavity mirrors by including a damping rate $\kappa$ in the Heisenberg equations of the field operators. Obviously, at the chosen low cavity decay rate of $\kappa = 1 \omega_r$ the effect of photon loss does not change the dynamics significantly. Note that already the initial state of the system $\ket{\Psi_0}$ has a nonzero overlap with the solution, so that we do not start at zero overlap.
To highlight the limits of the success rate for very long ramp times in more detail, we also plot the negative logarithm of the infidelity, i.e. the deviation of the fidelity from one, i.e., ($1-\mathcal{F}$) in \frefa{fig:nc3fid} (b). Somewhat surprisingly, already at about $\omega_r t_f=300$ the fidelity for the adiabatic Hamiltonian flattens and then even decreases, which could be attributed to the openness of the system. In contrast, the accuracy obtained from propagation of the full quantum Hamiltonian improves further and reaches a much higher maximum around a simulation time of $\omega_r t_f \approx 1300$. Again also in this case much longer simulation times do further reduce the fidelity, hinting for an ideal simulation time to maximize the fidelity. The reason behind this is not obvious and finding out in detail would probably need extensive simulations with higher numerical accuracy.

As shown e.g. in Ref. \cite{hauke2015}, there is a close connection between the atom-field entanglement at the end of the adiabatic sweep and the success probability, as any remaining entanglement would lead to a mixed state in atomic space after tracing out the field degrees of freedom. Hence, while, as we will see below, large entanglement during the annealing process is vital for success, it is detrimental if it remains at the end. To quantify the amount of entanglement in our system we use the entanglement entropy, i.e. the Von Neumann entropy of the reduced mode density matrix. We will thus concentrate at two specific properties here, namely the maximum entanglement present during a sweep and the remaining amount of entanglement at the end of the evolution, further on called the "final entanglement".\par

\begin{figure}[h!]
	\centering
	\includegraphics[scale=0.4]{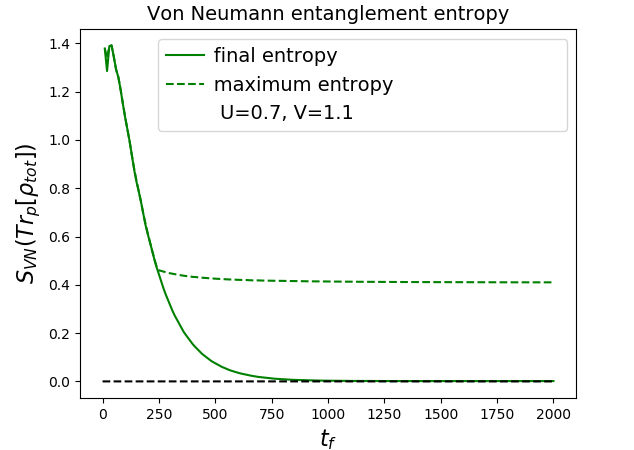}
	\caption{Maximum entanglement entropy attained during the sweep (dashed line) and the final entropy remaining after the sweep (solid line) as a function of total simulation time for parameters as in \frefa{fig:nc3fid}.}
	\label{fig:nc3ent}
\end{figure}
\frefa{fig:nc3ent} depicts these two key quantities as a function of the total simulation time $\omega_r t_f$. We see that with increasing simulation time the final entropy at the end of the evolution shrinks towards zero and the particles end up in the correct factorized state representing the sought solution. Note also the build up of a peak of maximum entanglement with simulation time until a certain optimal time. For longer time less entanglement builds up but it reaches a finite plateau at very long sweep times while the final entropy drops down to zero again. The splitting of the two values coincides with a peak in the process fidelity shown in \frefa{fig:nc3fid}. This suggests a close relation of success fidelity and the ability to get rid of the entanglement for a given simulation time.

\section{Annealing with semi-classical approximation}\label{sec:semi-class}

Let us now turn to the central question that we want to address in this work and compare the success of the annealing procedure in the quantum case as above with a semi-classical Hamiltonian (\ref{eq:H-semi-class}) based on mean fields to mediate the atom-atom interactions. Of course, for such a classical field description, no atom field entanglement can build up. In particular, we are looking for parameter regions, where the semi-classical model fails to reliably produce the desired solution, while the full quantum model still succeeds.\par

By evaluating the solution fidelity of the semi-classical model for a wide parameter range, we find a very strong dependence of the success fidelity on the particle-particle interaction parameter $U$ and the introduced impurity magnitude $V$. Only for small onsite interaction strength, the semi-classical approximation of the dynamics reliably finds the correct solution. Increasing interactions we see a surprisingly sudden jump of the success rate to low probability below $0.5$. Further increasing repulsive interaction, so that the final ground state energy only exhibits a very small gap to the first excitation, the classical model virtually never succeeds.

\begin{figure}[h!]
	\centering
	\includegraphics[width=0.45\textwidth]{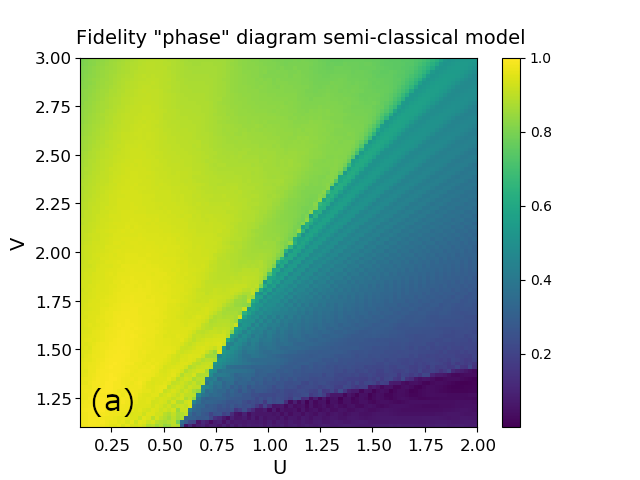} 
       \includegraphics[width=0.45\textwidth]{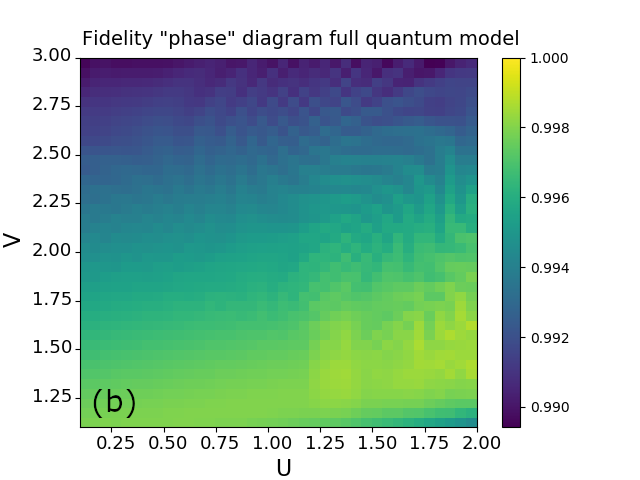}
	\caption{ (a) Color density plot (phase diagram) showing the solution fidelity after an adiabatic sweep using the Hamiltonian with classical mean-field approximation (\ref{eq:H-semi-class}) and as well as (b) for  the full quantum Hamiltonian as a function of the impurity strength $V$ along y and the particle interaction strength $U$ along x. Chosen parameters are: $t_f=1000$. For (a): $\Delta = -1$, $\tilde{J}=1$ and for (b): $\Delta = -5$, $\tilde{J}=\sqrt{5}$.}
	\label{fig:phase_diag}
\end{figure}

Hence, here we can identify a clear quantum improvement as shown in \frefa{fig:phase_diag} (a) the color density plot of the fidelity for a range of $V$ and $U$ parameters. The plot resembles a phase diagram a lot when we show the success fidelity of the final state using (\ref{eq:H-semi-class}). These results hint at a simple qualitative explanation. Since local particle-particle interaction favors systems where particles are separated, increasing $U$ reduces the wave-function amplitudes for the particles to occupy the same lattice site. However, in the energy spectrum of the system for corresponding values of $U$, we still find a gap from a de-localized state to the ground state with two particles occupying $\hat{n}_3$. Accordingly, the adiabatic sweep using the full quantum Hamiltonian (\ref{eq:H}) finds the correct solution with almost $99\%$ probability also for higher $U$ values as shown in \frefa{fig:phase_diag} (b). Note the different color scale range for the two plots differing by two orders of magnitude.

\begin{figure}[h]
  \centering
    \includegraphics[width= 0.45\columnwidth]{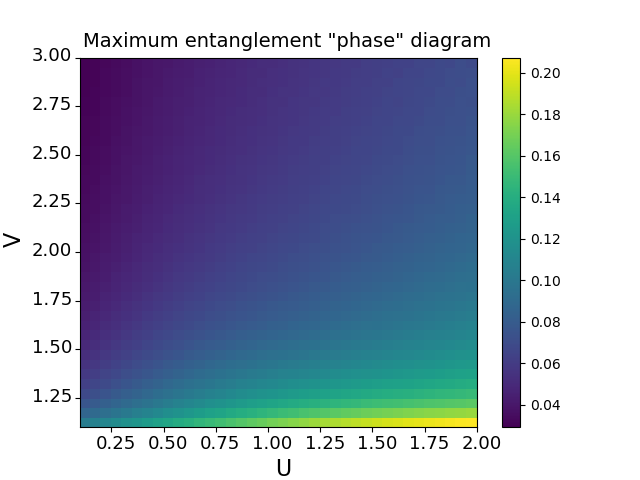}
    \caption{Maximum of the entanglement entropy attained for the full quantum model during a slow sweep as a function of the impurity strength $V$ on the y-axis and the particle interaction strength $U$ along x. Chosen parameters are: $t_f=1000$, $\Delta = -5$, $\tilde{J}=\sqrt{5}$.}
    \label{fig:Entanglement_t}
\end{figure}

{\bf Entanglement and Fidelity:} It is, of course, important to compare this behavior to the appearance of atom-field entanglement during the time evolution of the full quantum dynamics. While the final entanglement almost vanishes at the end, we still find a strong maximum during evolution as shown in \frefa{fig:Entanglement_t} for the same parameters as above.
Studying the plots side by side reveals a correlation between regions of high entanglement during quantum annealing and a very low fidelity of the semi-classical solution. Especially in the right lower corner, where the classical fidelity is close to zero, we observe a peak in the entanglement, which continues along the x-axis but reduces in intensity. One can faintly also recognize a similar triangular area of the high entanglement region. 

Obviously, regions of high fidelity are correlated to regions of low entanglement entropy and vaguely exhibit the same transition line. In particular in low fidelity - high entropy regions the indication of a direct connection between entanglement and the success probability for the semi-classical model look convincing although the transition lines are not as clear cut. As we have discussed above a classical field approximation removes the possibility of entangled particle-mode states. Hence, it is not surprising that the discrepancy between the success probabilities for the full quantum model and the semi-classical model is connected to the entanglement. Interestingly, we can also see a clear correspondence between the entanglement peak and the reduced fidelity in the lower right corner of \frefa{fig:phase_diag} (b). Although the entanglement can be quantitatively small, the maximum entanglement shows a clear correlation with the solution fidelity of the full quantum system. 

\section{Speedup of numerical simulations by a lower photon number cut-off}\label{sec:cut-off}
\label{chap:cut-off}
The above considerations strongly support the idea that a real quantum advantage could be found in annealing processes. Of course, numerically we cannot find the exact solution for the quantum evolution, as we always need to truncate the simulation Hilbert space and in particular the maximum photon number allowed in the modes.
However, in our simple model we can still use a sufficiently high cutoff to make the difference to the full solution negligibly small. Surprisingly, it seems that the errors introduced by a lower and even nonphysically low cutoff often are not detrimental, but even help to find a higher fidelity solution. This comes in addition to the numerical speedup gained from a smaller computational basis. Does this mean that there is no quantum advantage in real life after all? Can there be a better model than quantum mechanics to do annealing ?  

 In this final section we will study this intriguing phenomenon in more detail and work to quantify the influence of photon number cutoff and simulation time on the success probability.

\begin{figure}[h!]
	\centering 
	\includegraphics[width = 0.2\textwidth]{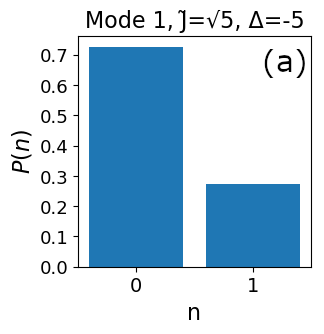}
	\includegraphics[width = 0.2\textwidth]{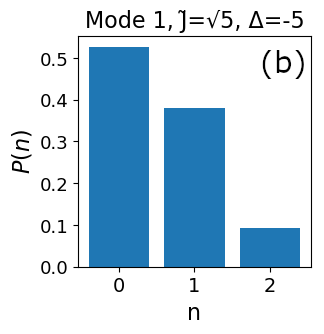}
	\includegraphics[width = 0.2\textwidth]{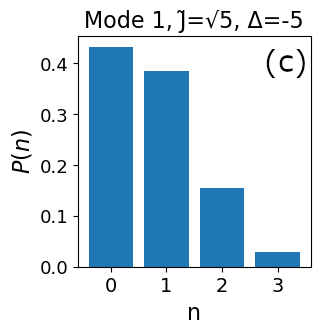}
	\includegraphics[width = 0.2\textwidth]{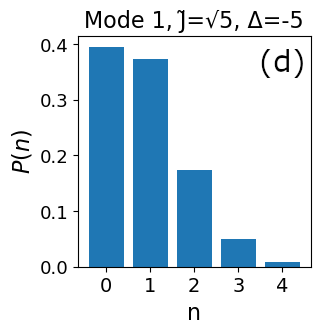}
	\caption{Photon number occupation probabilities for one mode at the end of the adiabatic evolution for different cut-offs of the mode Hilbert space  $nc= 1$ (a), $nc=2$ (b), $nc=3$ (c) and $nc=4$ (d) from left to right.}
	\label{fig:occ_nc}
\end{figure}

As we argued above, the photon number distributions obtained in the simulations suggest that a cutoff at $nc=3$ photons leads to a very good accuracy of the numerical solution at a tractable numerical complexity. It allows to reliably describe the field quantum state during the whole evolution time, missing only higher occupations well below a percent and no visible changes in the dynamics for higher cutoffs ($nc=4$).

Nevertheless one can hope that a lower cutoff, which speeds up the calculations, is still good enough to obtain useful results in shorter time. To our surprise it turns out, that not everything behaves as anticipated here. Let us take a closer look at simulations based on lower photon number cut-offs, which clearly reduce the quality of the numerical solution of the Schrödinger equation. 

\begin{figure}[h!]
	\centering 
	\includegraphics[width = 0.35\textwidth]{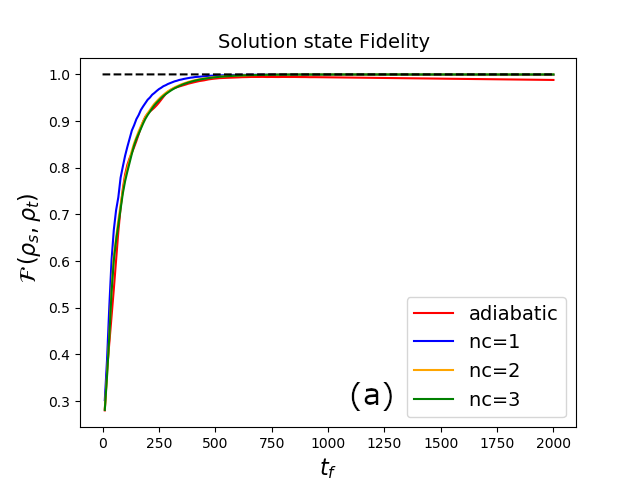}
	\includegraphics[width = 0.35\textwidth]{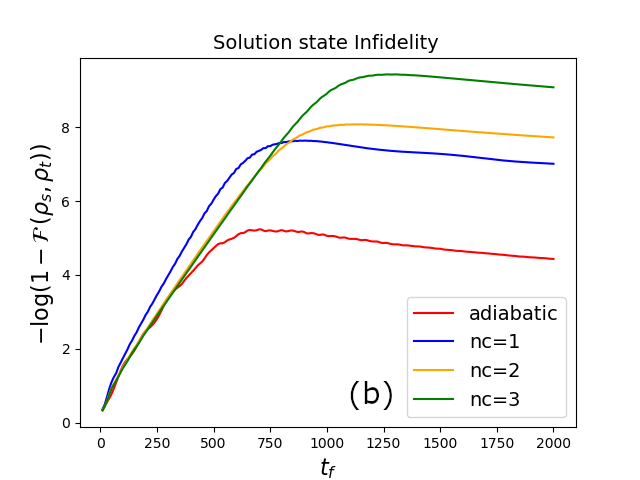}
	\caption{(a) Solution fidelity between final state of the adiabatic evolution and target state as a function of sweep duration $t_f$ for different photon number cut-offs (blue, yellow, green) and the adiabatic Hamiltonian (red).  (b) The same result shown via the negative logarithm of the infidelity demonstrate a clear improvement for higher cutoffs only for a sufficiently long sweep duration. Parameters are: $J=0.1$, $U=0.7$, $V=1.1$, $\tilde{J}^2=5=\Delta$ and $\kappa=1$.}
	\label{fig:cut-off}
\end{figure}

Following the annealing procedure as above, in \frefa{fig:occ_nc}. we plot the final state photon number occupation probabilities for different cut-offs $nc\in\set{1,2,3}$, from left to right. It is sufficient to look at one mode here as the second one does not differ significantly. Of course with growing cut-off the vacuum contribution is decreasing, which clearly proves that the solutions obtained for lower cut-off are not faithful approximation to the full quantum dynamics. Nevertheless, as shown in \frefa{fig:cut-off} (a) the success probability for finding the desired atomic final is virtually the same in all case. Actually for fast ramps the lowest cutoff $nc=1$ simulations, where the modes are essentially approximated by q-bits, lead to an even higher success probability than higher cut-offs. Note that a semi-classical approximation completely fails here and some quantumness is still needed.    

Let us zoom in a bit and plot the infidelity for the three different cut-offs on a logarithmic scale. This confirms that the numerically not justifiable nc=1 cutoff leads to the best results for fast sweeps. Hence we can win in several respects by not reliably solving for the full Schrödinger dynamics. We can use a smaller Hilbert space, which saves memory and computation time as well as a faster ramp time, which again speeds up the computation. Of course, it is hard to envisage how this quantum inspired algorithm \cite{heim2015} could be implemented in an experiment.

Fortunately a second look at \frefa{fig:cut-off} (b) reveals a second interesting feature. When we implement a careful and numerically reliable solution using a slow sweep we find the highest single run fidelity. Quantitatively a larger Hilbert space can reduce minimal possible infidelity by even a couple orders of magnitude at the expense of a much slower computation. While this might seem not so useful at the end, the picture could strongly change in much more complex setups, where much lower single run success is expected and only the full dynamics will yield useful performance.   

\begin{figure}[h!]
	\centering
	\includegraphics[width = 0.35\textwidth]{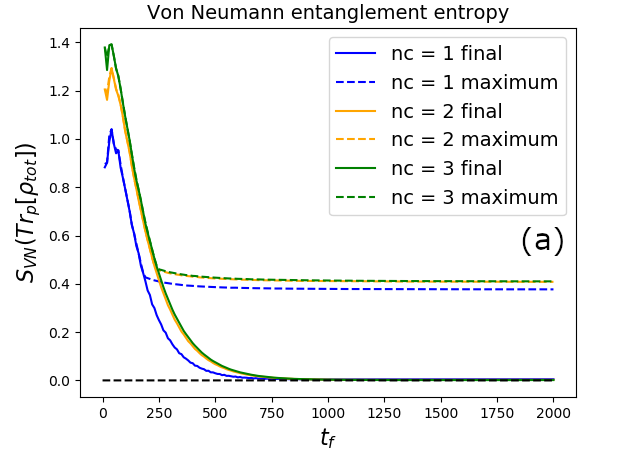}
	\includegraphics[width = 0.35\textwidth]{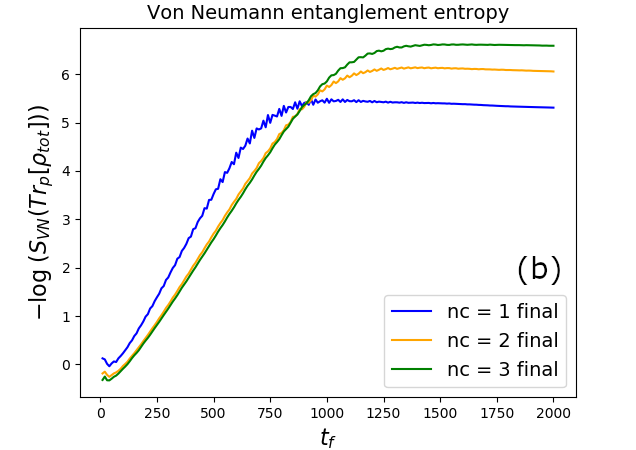}
	\caption{ (a) Maximal (dashed lines) and final (solid lines) generated atom-field entanglement as a function of simulation time $t_f$  for the different photon number cut-offs as above. (b) The same results shown on a negative logarithmic scale exhibit the diminishing final entanglement responsible for the higher process success rate at larger cutoff and slower sweeps. Same parameters as in \frefa{fig:cut-off}.}
	\label{fig:cut-off_ent}
\end{figure}

What else can we learn from this example? As discussed above the atom-field entanglement remaining at the end of the adiabatic evolution interferes with the ability to find a good solution. In \frefa{fig:cut-off_ent} one finds a clear signature of this effect comparing the final states for different photon number cut-offs. As the Hilbert space is larger, higher cut-offs allow for the buildup of more maximal entanglement in the process, which would suggest a better optimization performance. However, for relatively short simulation times, the final entanglement reduces towards zero at the end much faster for lower cutoff than for the higher cut-offs. Hence for short times the penalty of more remaining final entanglement for the full cutoff dominates over the gain from a higher in process entanglement. This fits well to the behavior of the fidelity in \frefa{fig:cut-off}. The logarithmic plot on the right side reproduces the identical behavior for long simulation times. There, the lowest cut-off $nc=1$ reaches a steady state and for the highest cut-off $nc=3$ long ramp times allow to minimize the final entanglement the most. 

\section{Conclusions}
In this work we identified and studied a generic toy system, where a clear quantitative improvement of the annealing process for a full quantum simulation versus the semi-classical counterpart can be demonstrated. As the two approaches only differ in the representation of the field modes mediating interactions by classical or quantum harmonic oscillators, the beneficial role of quantum dynamics is very clearly revealed. In this sense, we directly compare solutions of the Schrödinger equation and corresponding Maxwell equations. While superposition and interference of field mode amplitudes is possible in both cases, only the quantum description opens the possibility of entanglement, which manifests itself as the key ingredient towards generating a quantum advantage. Although, an in depth discussion in reference to the debate around the \emph{quantumness} of the D-Wave quantum annealer \cite{smolin2014a,wang2013a} is beyond this work, let us emphasis once more the fact that in our comparison of quantum annealing and a semi-classical counterpart, the classicality of the system solely arises from considering classical field modes and hence eliminating the possibility for light-matter entanglement, therefore enabling a clearer investigation into the quantumness of the system.

However, there is not a quantitative one to one correspondence between entanglement and improvement. On the one hand we show a direct connection between the maximal atom-field entanglement in the dynamics and a high probability to find the optimal solution at the end. On the other hand, for a too fast sweep time, the final atomic state keeps some entanglement with the light, which diminishes the final solution fidelity. As the desired final ground state is separable, a strong quantitative anti-correlation between the success probability and final state entanglement is quite obvious. In summary, highly reliable annealing needs strong entanglement during the annealing process but vanishing entanglement at the end.   

This feature can help to explain the surprising improvement one gets from using photon number cut-offs too low to accurately represent the full quantum dynamics in the simulations. This unphysical cutoff prevents the buildup of large entanglement at short simulated times, so that only very little entanglement remains at the end of the sweep. For certain operating parameters, the benefit of the final disentanglement turns out to be more important than the extra fidelity gain one obtains from a larger maximum entanglement during the process. Nevertheless we see that the limitation of the Hilbert space and thus maximal entanglement also limits the maximal obtainable fidelity even for very long simulations times. We did not find, however, any regime where a classical description gives the best results. 

Of course, to argue for a possible advantage of quantum annealing in general, there is still quite some distance to go. At first, a deeper investigation into the ideal amount of entanglement for finding an optimal solution is vital for identifying the optimal operation parameters. The central question here is the scaling of optimal entanglement with system size and whether this stays essential for large systems. Finally, also the requirement of disentanglement at the end needs to be shown not to scale unfavorably with system size. While all of these assumptions look reasonable, the corresponding substantial proofs are challenging and an experimental evaluation might be a more reasonable approach here. While for implementing quantum inspired numerical algorithms \cite{heim2015} a unfaithful representation of the field Hilbert space can speed up calculations and increase the solution fidelity at the same time, it is unclear how this could be mimicked in experimental devices.           

As said above, the central question to be answered is the scaling of the fidelity improvement through entanglement in larger systems. Some preliminary results obtained in simulations of the N-Queens problem hint for the growing importance of entanglement with system size \cite{torggler2019}. Here numerical studies based on the Heisenberg equations of the system operators using a so-called cumulant expansion \cite{heim2015} should be a promising method to better uncover the role of entanglement in the annealing dynamics in larger systems. To study the ultimate performance of an open quantum system approach, the effect of photon decay through the cavity mirrors needs to be systematically included in form of a corresponding master equation approach. Fortunately, using proper operating parameters in open systems, intrinsic cooling could be used to compensate for non-adiabatic errors and allow for a simulation speedup. Hence there are ample options and possibilities for future studies in various interesting directions.        

\ack
We thank Valentin Torggler and Wolfgang Lechner for valuable discussions. The numerical simulations were performed with the open-source framework QuantumOptics.jl~\cite{kramer2018} and the graphs were produced with the open-source library Matplotlib~\cite{hunter2007}. H.R. acknowledges support from the Austrian Science Fund under project P29318-N27. 

\section*{References}


\begin{thebibliography}{10}
	\expandafter\ifx\csname url\endcsname\relax
	\def\url#1{{\tt #1}}\fi
	\expandafter\ifx\csname urlprefix\endcsname\relax\def\urlprefix{URL }\fi
	\providecommand{\eprint}[2][]{\url{#2}}
	
	\bibitem{benioff1980}
	Benioff P 1980 {\em Journal of Statistical Physics\/} {\bf 22} 563--591
	
	\bibitem{feynman1982}
	Feynman R~P 1982 {\em International Journal of Theoretical Physics\/} {\bf 21}
	467--488
	
	\bibitem{deutsch1985}
	Deutsch D 1985 {\em Proceedings of the Royal Society of London Series A\/} {\bf
		400} 97--117
	
	\bibitem{dowling2002}
	Dowling J~P and Milburn G~J 2002 {\em Quantum {{Technology}}: {{The Second
				Quantum Revolution}}\/}
	
	\bibitem{das2005}
	Das A and Chakrabarti B~K 2005 {\em Quantum Annealing and Related Optimization
		Methods\/} vol 679 ({Springer Science \& Business Media})
	
	\bibitem{ray1989}
	Ray P, Chakrabarti B~K and Chakrabarti A 1989 {\em Physical Review B\/} {\bf
		39} 11828--11832
	
	\bibitem{apolloni1989}
	Apolloni B, Carvalho C and de~Falco D 1989 {\em Stochastic Processes and their
		Applications\/} {\bf 33} 233--244 ISSN 0304-4149
	
	\bibitem{apolloni1990}
	Apolloni B, {Cesa-Bianchi} N and {de Falco} D 1990 A numerical implementation
	of ``quantum annealing'' {\em Stochastic Processes, Physics and Geometry
		({{Ascona}} and {{Locarno}}, 1988)\/} ({World Sci. Publ., Teaneck, NJ}) pp
	97--111
	
	\bibitem{kadowaki1998}
	Kadowaki T and Nishimori H 1998 {\em Physical Review E\/} {\bf 58} 5355--5363
	
	\bibitem{born1928}
	Born M and Fock V 1928 {\em Zeitschrift f\"ur Physik\/} {\bf 51} 165--180 ISSN
	0044-3328
	
	\bibitem{britton2012}
	Britton J~W, Sawyer B~C, Keith A~C, Wang C~C~J, Freericks J~K, Uys H, Biercuk
	M~J and Bollinger J~J 2012 {\em Nature\/} {\bf 484} 489--492 ISSN 1476-4687
	
	\bibitem{bunyk2014}
	Bunyk P~I, Hoskinson E~M, Johnson M~W, Tolkacheva E, Altomare F, Berkley A~J,
	Harris R, Hilton J~P, Lanting T, Przybysz A~J and {al} e 2014 {\em IEEE
		Transactions on Applied Superconductivity\/} {\bf 24} 1--10 ISSN 1558-2515
	
	\bibitem{jotzu2014}
	Jotzu G, Messer M, Desbuquois R, Lebrat M, Uehlinger T, Greif D and Esslinger T
	2014 {\em Nature\/} {\bf 515} 237--240 ISSN 1476-4687
	
	\bibitem{johnson2011}
	Johnson M~W, Amin M~H~S, Gildert S, Lanting T, Hamze F, Dickson N, Harris R,
	Berkley A~J, Johansson J, Bunyk P, Chapple E~M, Enderud C, Hilton J~P, Karimi
	K, Ladizinsky E, Ladizinsky N, Oh T, Perminov I, Rich C, Thom M~C, Tolkacheva
	E, Truncik C~J~S, Uchaikin S, Wang J, Wilson B and Rose G 2011 {\em Nature\/}
	{\bf 473} 194--198 ISSN 1476-4687
	
	\bibitem{hauke2020}
	Hauke P, Katzgraber H~G, Lechner W, Nishimori H and Oliver W~D 2020 {\em
		Reports on Progress in Physics\/} {\bf 83} 054401 ISSN 1361-6633
	
	\bibitem{heim2015}
	Heim B, R{\o}nnow T~F, Isakov S~V and Troyer M 2015 {\em Science\/} {\bf 348}
	215--217
	
	\bibitem{arrazola2019}
	Arrazola J~M, Delgado A, Bardhan B~R and Lloyd S 2019 {\em arXiv preprint
		arXiv:1905.10415\/} (\textit{Preprint} \eprint{1905.10415})
	
	\bibitem{das2008}
	Das A and Chakrabarti B~K 2008 {\em Reviews of Modern Physics\/} {\bf 80}
	1061--1081
	
	\bibitem{albash2018a}
	Albash T and Lidar D~A 2018 {\em Reviews of Modern Physics\/} {\bf 90} 015002
	
	\bibitem{torggler2017}
	Torggler V, Kr{\"a}mer S and Ritsch H 2017 {\em Physical Review A\/} {\bf 95}
	032310
	
	\bibitem{letavec2002}
	Letavec C and Ruggiero J 2002 {\em INFORMS Transactions on Education\/} {\bf 2}
	101--103
	
	\bibitem{torggler2019}
	Torggler V, Aumann P, Ritsch H and Lechner W 2019 {\em Quantum\/} {\bf 3} 149
	ISSN 2521-327X
	
	\bibitem{gent2017}
	Gent I, Jefferson C and Nightingale P 2017 {\em Journal of Artificial
		Intelligence Research (JAIR)\/}
	
	\bibitem{dagotto2005complexity}
	Dagotto E 2005 {\em Science\/} {\bf 309} 257--262
	
	\bibitem{mekhov2012}
	Mekhov I~B and Ritsch H 2012 {\em Journal of Physics B: Atomic, Molecular and
		Optical Physics\/} {\bf 45} 102001
	
	\bibitem{maschler2008}
	Maschler C, Mekhov I~B and Ritsch H 2008 {\em The European Physical Journal
		D\/} {\bf 46} 545--560 ISSN 1434-6079
	
	\bibitem{torggler2014}
	Torggler V and Ritsch H 2014 {\em Optica\/} {\bf 1} 336--342
	
	\bibitem{keller2018}
	Keller T, Torggler V, J{\"a}ger S~B, Sch{\"u}tz S, Ritsch H and Morigi G 2018
	{\em New Journal of Physics\/} {\bf 20} 025004
	
	\bibitem{kramer2014}
	Kr{\"a}mer S and Ritsch H 2014 {\em Physical Review A\/} {\bf 90} 033833
	
	\bibitem{breuer2002}
	Breuer H, Breuer P, Petruccione F and Petruccione S 2002 {\em The {{Theory}} of
		{{Open Quantum Systems}}\/} ({Oxford University Press}) ISBN
	978-0-19-852063-4
	
	\bibitem{hauke2015}
	Hauke P, Bonnes L, Heyl M and Lechner W 2015 {\em Frontiers in Physics\/} {\bf
		3} 21 ISSN 2296-424X
	
	\bibitem{smolin2014a}
	Smolin J~A and Smith G 2014 {\em Frontiers in Physics\/} {\bf 2} ISSN 2296-424X
	
	\bibitem{wang2013a}
	Wang L, R{\o}nnow T~F, Boixo S, Isakov S~V, Wang Z, Wecker D, Lidar D~A,
	Martinis J~M and Troyer M 2013 {\em arXiv:1305.5837 [quant-ph]\/}
	(\textit{Preprint} \eprint{1305.5837})
	
	\bibitem{kramer2018}
	Kr{\"a}mer S, Plankensteiner D, Ostermann L and Ritsch H 2018 {\em Computer
		Physics Communications\/} {\bf 227} 109--116
	
	\bibitem{hunter2007}
	Hunter J~D 2007 {\em Comput. Sci. Eng.\/} {\bf 9} 90--95
	
\end{thebibliography}
\providecommand{\newblock}{}

\end{document}